\documentclass[]{spie}  %>>> use for US letter paper
\usepackage{amsmath,amsfonts,amssymb}
\usepackage{graphicx}
\usepackage[colorlinks=true, allcolors=blue]{hyperref}
\usepackage{bm}

\def\A{\mathcal{A}}

\def\ii{\mathrm{i}}

\def\br{\bm{\rho}}

\everymath{\displaystyle}

\title{Plenoptic microscopy and photography from intensity correlations}

\author[a,b]{Francesco V. Pepe}
\author[b]{Francesco Di Lena}
\author[a,b]{Augusto Garuccio}
\author[a,b]{Davide Giannella}
\author[c]{Alessandro Lupo}
\author[a]{Gianlorenzo Massaro}
\author[a]{Alessio Scagliola}
\author[a,b]{Francesco Scattarella}
\author[b]{\\ Sergii Vasiukov} 
\author[a,b]{Milena D'Angelo}

\affil[a]{Dipartimento Interateneo di Fisica, Università degli studi di Bari, I-70126 Bari, Italy}
\affil[b]{INFN, Sezione di Bari, I-70125 Bari, Italy}
\affil[c]{Laboratoire d'Information Quantique, CP 224, Universitè libre de Bruxelles, Av. F. D. Roosevelt 50, B-1050, Bruxelles, Belgium}

\authorinfo{Send correspondence to Francesco V. Pepe. E-mail: francesco.pepe@ba.infn.it}

\makeatother

\begin{document}
\maketitle
\begin{abstract}
We present novel methods to perform plenoptic imaging at the diffraction limit by measuring intensity correlations of light. The first method is oriented towards plenoptic microscopy, a promising technique which allows refocusing and depth-of-field enhancement, in post-processing, as well as scanning free 3D imaging. To overcome the limitations of standard plenoptic microscopes, we propose an adaptation of Correlation Plenoptic Imaging (CPI) to the working conditions of microscopy. We consider and compare different architectures of CPI microscopes, and discuss the improved robustness with respect to previous protocols against turbulence around the sample. The second method is based on measuring correlations between the images of two reference planes, arbitrarily chosen within the tridimensional scene of interest, providing an unprecedented combination of image resolution and depth of field. The results lead the way towards the realization of compact designs for CPI devices.
\end{abstract}


\section{INTRODUCTION}
\label{sec:intro} 

Plenoptic imaging (PI) is an optical imaging technique that enables to measure the distribution of the 3D light field with a single sensor exposure. The acquired information encodes at once both the spatial intensity distribution
of light in a given plane, which is available even when performing
standard imaging, and the direction of light propagating through that plane\cite{PI,PI1}. The combination of spatial and directional information makes it possible to reconstruct light paths in post-processing, unfolding a set of interesting possibilities such as refocusing, depth-of-field (DOF) extension, variation of the point of view on the scene. To date, PI is also one of the most convenient methods to obtain scanning-free single-shot 3D reconstruction\cite{Broxton:13,Xiao:13,PI6,c72b58c6767e455488b7dd420d55b06a}. 

In its conventional realization, a plenoptic imaging device collects both spatial and directional information on the same sensor. A microlens array is inserted between the detector and an otherwise standard optical apparatus\cite{PI2,PI3,PI4}. The presence of microlenses, albeit necessary to encode the directional of light on the sensor, drastically reduces the spatial resolution of the system\cite{5989827}. This translates into a marked trade-off
between DOF and resolution, that can be mitigated only by applying signal reconstruction techniques\cite{PI5,PI6,PI7}. An alternative technique called \textit{correlation plenoptic imaging} (CPI) was introduced\cite{firstCPI} in order to overcoming as much as possible the aforementioed tradeoff, thus enabling diffraction-limited imaging in a plenoptic apparatus. The main idea on which CPI is based is to decouple spatial and directional information, that determine the resolution and DOF properties of the plenoptic image,
by performing correlation measurements measurements between the intensities registered by two separate sensors. Dividing the two measurements onto two distinct detectors allows not to sacrifice any
longer one in favor of the other\cite{PhysRevLett.119.243602}. The working principle of CPI is valid for different statistical properties of the detected light: in particular, schemes based on the spatio-temporal correlations
of entangled photon pairs\cite{CPIent,overview} and chaotic
light\cite{overview,article} have been developed.

In this paper, we outline two recently developed methods to perform plenoptic imaging at the diffraction limit by measuring intensity correlations of light. The first method, called \textit{correlation plenoptic microscopy} (CPM)\cite{CPM}, represents an adaptation of CPI to the typical working conditions of microscopic imaging, in which scattering or fluorescent samples, often surrounded by diffusive effects, are examined. We consider and compare different architectures of CPI microscopes, and discuss the improved robustness with respect to previous protocols against turbulence around the sample, highlight the physical limits of the described technique. The second method, called \textit{correlation plenoptic imaging between arbitrary planes} (CPI-AP)\cite{CPI-AP}, is based on measuring correlations between the images of two reference planes, arbitrarily chosen within the 3D scene of interest. This protocol enables to achieve an unprecedented combination of image resolution and depth of field. Moreover, due to the simple structure of the setup, CPI-AP leads the way towards the development of compact designs for correlation plenoptic imaging devices based on chaotic light, that can be considered an enhanced version of standard photography cameras.

\section{Correlation plenoptic microscopy}\label{sec:working}

\begin{figure}
\centering
\includegraphics[width=0.7\textwidth]{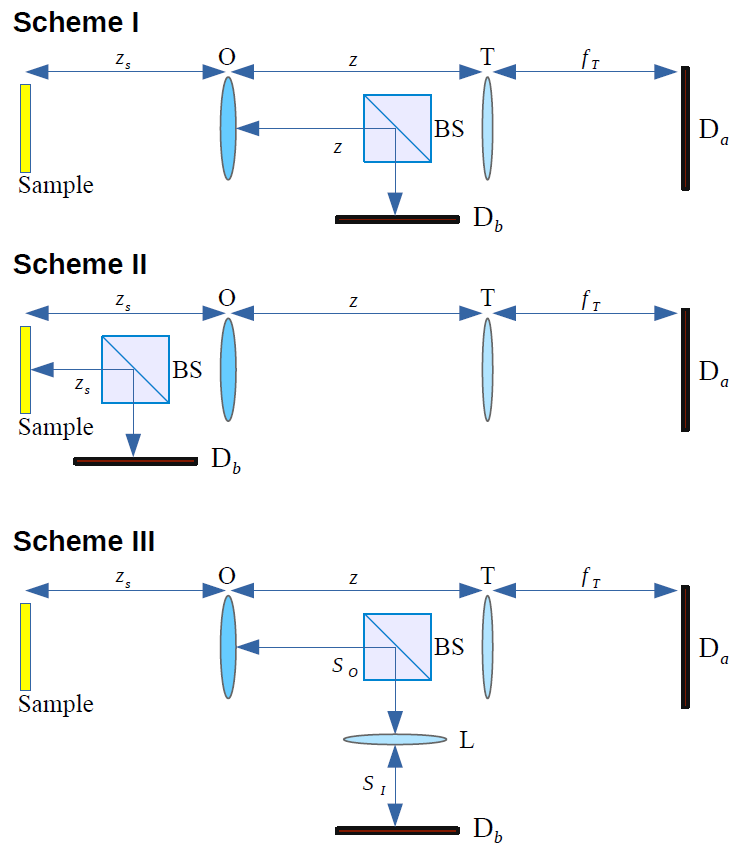}
\caption{Schematic representation of three possible realizations of Correlation Plenoptic Microscopy (CPM). All of them feature a sample, modelled as a chaotic light emitter, an objective lens (O) of focal length $f_O$, a tube lens (T) of focal length $f_T$, a beam splitter (BS) and two high-resolution sensors $\mathrm{D}_a$ and $\mathrm{D}_b$; the schemes differ by the role played by $\mathrm{D}_b$ in retrieving the required directional information on light propagation. In all the schemes, the lenses O and T are at a distance $z$ from each other, the detector $\mathrm{D}_a$ is in the focal plane of the tube lens T, and the sample is at a generic distance $z_s$ from the objective lens O. In protocols I and II, $\mathrm{D}_b$ is at the same distance $z+z_S$ and $z_S$, respectively, from the source. In protocol III, $\mathrm{D}_b$ is in the conjugate plane of the lens O, as defined by the auxiliary lens $L$ of focal length $f_L$.}\label{fig:CPM_setups}
\end{figure}

CPI was originally conceived as an extension of ghost imaging\cite{pittman,ghost,gatti,laserphys,valencia,scarcelliPRL} to accomplish the tasks of plenoptic imaging. As in the forerunner technique, in the first CPI schemes an object was illuminated by only one of the two beams separated by a beam splitter\cite{overview}. This configuration represented a strong obstruction to the application of CPI in contexts where either the object diffusive or it can be modelled as a chaotic source itself (e.g., it is fluorescent). Actually, the working principle of CPI entails that correlations originated from the illuminating source must be preserved in both paths between the source and the sensors. The described scheme is highly sensitive to turbulence and scattering in the surrounding of the object. This issue was addressed by developing CPM\cite{CPM}. As shown in Fig.~\ref{fig:CPM_setups}, by considering novel CPI schemes where the sample is placed in any case before the beam splitter (BS) and can thus be a generic transmissive, reflective, scattering, or even emitting object. The CPM approach also increases the robustness against turbulence.

We consider the case in which the conventional image of the a sample (represented as a planar object in Fig.~\ref{fig:CPM_setups}, for the sake of simplicity) is mapped on the high resolution sensor $\mathrm{D}_a$ by a standard microscope, composed of an objective lens (O) and a tube lens (T). The sample is placed at a distance $z_s$ from O (more precisely, from its first principal plane). Since $\mathrm{D}_a$ lies in the back focal plane of the tube lens, the image of the sample is focused when it is placed in the front focal plane of the objective lens, namely, $z_s=f_O$, magnified by an absolute factor $M=f_T/f_O$. Plenoptic imaging is obtained by measuring correlations between $\mathrm{D}_a$ and another high resolution sensor $\mathrm{D}_b$, placed in the beam reflected off the beam splitter, and aimed at imaging one of the microscope lenses. The pixel-by-pixel correlation of the intensity fluctuations is evaluated to retrieve the \textit{plenoptic image}, encoded in the correlation function
\begin{equation}\label{Gamma}
\Gamma(\bm{\rho}_a,\bm{\rho}_b) = \langle \Delta I_a (\br_a) \Delta I_b (\br_b) \rangle,
\end{equation}
where $\langle\dots\rangle$ denotes an average on the chaotic statistics of light from the sample, $I_{a,b}(\br_{a,b})$ are the intensities at positions $\br_a$ and $\br_b$ on each detector plane, and $\langle I_{a,b}\rangle$ their statistical averages. The schemes actually differ only by the strategy to acquire information on the direction of light:
\begin{itemize}
    \item in Scheme I, a ``ghost image'' of the lens T is mapped on the sensor $\mathrm{D}_b$;
    \item in Scheme II, a ``ghost image'' of the lens O is mapped on the sensor $\mathrm{D}_b$;
    \item in Scheme III, a standard image of the lens T is mapped by the additional lens L on the sensor $\mathrm{D}_b$.
\end{itemize}
Here, we will review the main properties of Scheme III, in which an image of the objective is provided as in Scheme II, while keeping the beam splitter between the objective and the tube lens, as Scheme I. The image of the objective is reproduced by a lens L, of focal length $f_L$ on the sensor array $\mathrm{D}_b$. The distances $S_O$ and $S_I$ satisfy the usual thin lens equation: $1/S_I+1/S_O=1/f_L$. The focusing of the objective on $\mathrm{D}_b$ can thus be realized by direct observation on $\mathrm{D}_b$, and no delicate fine tuning along path $b$ is required.

To account for the partial coherence of light on the sample, which can be a relevant effect in microscopy, we assume that the equal-time correlators between arbitrary field components $V(\br_s)$, evaluated on the sample surface, are well described by a Shell model \cite{mandel}:
\begin{equation}
\left\langle V(\br_s) V^*(\br_s') \right\rangle = A(\br_s) A^*(\br_s') \exp\left(-\frac{(\br_s-\br_s')^2}{2\sigma_g^2}\right),
\end{equation}
where $A(\br_s)=\left\langle V(\br_s) \right\rangle$ is the amplitude profile and $\sigma_g$ is the transverse coherence length. The correlation between intensity fluctuations of light detected by the two sensors then reads\cite{CPM}
\begin{equation}\label{Gammageneral}
\Gamma(\bm{\rho}_a,\bm{\rho}_b) = \Biggl| \int d^2\bm{\rho}_s \int  d^2\bm{\rho}'_s \, g_a(\bm{\rho}_a,\bm{\rho}_s) g_b^*(\bm{\rho}_b,\bm{\rho}_s')  A(\br_s)A(\br_s') e^{-\frac{(\br_s-\br_s')^2}{2\sigma_g^2}} \Biggr|^2 ,
\end{equation}
with $g_a$ and $g_b$ the optical transfer functions describing paths $a$ and $b$ joining the sample with the detectors $\mathrm{D}_a$ and $\mathrm{D}_b$, respectively. In the limit of negligible transverse coherence, the correlation function in Eq.~\eqref{Gammageneral} depends only on the intensity profile $|A|^2$ of light on the sample plane. Such a feature improves robustness of the result against diffusive effects, that do not affect neither imaging and directional detection, as far as they occur on the sample. Such robustness, that is typical of incoherent imaging techniques, is not shared by previously developed CPI setups \cite{overview}, in which the directional reconstruction crucially depends on both the intensity and the phase of light on the object plane, making tasks like refocusing and 3D imaging highly sensitive to turbulence and scattering on the sample plane. In CPM, turbulence and scattering in the surrounding of the sample are also negligible within distances short enough not to affect the coherence of light, namely up to a distance  $d_t\ll L_t \delta/\lambda$, where $\lambda$ is the light wavelength, $\delta$ is the smallest length scale of details in the intensity profile of the source, and phase diffusion is characterized by a length scale of transverse spatial variations $L_t$.

The computation of $\Gamma(\br_a,\br_b)$ leads to the result\cite{CPM}
\begin{equation}\label{Gamma3}
\Gamma(\br_a,\br_b) = \mathcal{K} \left| \int d^2\br_s\int d^2\br_s'\int d^2\br_O\, A(\br_s)A^*(\br_s') 
P_O(\br_O) e^{\ii k \varphi_3(\br_s,\br_s'\br_O,\br_a,\br_b) - \frac{(\br_s-\br_s')^2}{2\sigma_g^2}} \right|^2
\end{equation}
with
\begin{equation}\label{phase3}
\varphi_3(\br_s,\br_s', \br_O,\br_a,\br_b) = \frac{1}{2 z_S} (\br_s^2-\br_s'^2) - \frac{\br_b \cdot \br_s'}{M_L z_S}  + \frac{1}{2} \left( \frac{1}{z_S} - \frac{1}{f_O} \right) \br_O^2 - \br_O \cdot \left( \frac{\br_a}{f_T} + \frac{\br_s}{z_S} \right) .
\end{equation}
where $M_L=S_I/S_O$ is the magnification of the image of the objective aperture $P_O$ on the sensor $\mathrm{D}_b$, and $\mathcal{K}$ is an irrelevant constant. In the case of focused sample ($z_S=f_O$), the integral over the directional detector $\mathrm{D}_b$ reduces to
\begin{equation}\label{focused2}
\Sigma(\br_a) = \int d^2 \br_b \, \left. \Gamma(\br_a,\br_b) \right|_{z_S=f_O} = \tilde{\mathcal{K}} \int d^2\br_s |A(\br_s)|^4 \left| \int d^2\br_O \, P_O(\br_O) \exp\left( - \frac{\ii k}{f_O} \br_O \cdot \left( \br_s + \frac{\br_a}{M} \right) \right) \right|^2,
\end{equation}
essentially encoding the focused image of the sensor, characterized by the same point-spread function (notice, however, that the intensity profile $|A|^2$ on the sample is squared, unlike in the first-order image). In the out-of-focus case, in the geometrical optics approximation and the assumption of negligible transverse coherence lead to the asymptotic result
\begin{equation}
\Gamma(\br_a,\br_b) \sim \left| P_O\left(-\frac{\br_b}{M_L}\right) \right|^4 \left|A \left( - \frac{z_S}{f_O} \frac{\br_a}{M} - \left( 1 - \frac{z_S}{f_O} \right) \frac{\br_b}{M_L} \right) \right|^4 .
\end{equation}
Therefore, in the geometrical optics regime, it appears that multiple images of the object are retrieved, each corresponding to a value of the coordinate $\br_b$ on the detector $\mathrm{D}_b$. However, directly performing an integration on directions (namely, on $\br_b$), as in \eqref{focused2}, leads to a blurred result, since sub-images with different alignments are summed together, unless $z_s$ is not farther from $f_O$ than the depth of field determined by the point-spread function. To realign sub-images, a refocusing procedure, analogous to the one applied in conventional plenoptic imaging, should be applied, to construct the function
\begin{equation}\label{refocus3}
\Gamma_{\mathrm{ref}}(\br_a,\br_b) = \Gamma \left[ \frac{f_O}{z_S} \br_a + \left( 1- \frac{f_O}{z_S} \right) \frac{M}{M_L} \br_b , \br_b \right] \sim
 \left| P_O\left(-\frac{\br_b}{M_L}\right) \right|^4 \left|A \left( - \frac{\br_a}{M} \right) \right|^4 .
\end{equation}
The decoupling between the image of the object and of the lens achieved in \eqref{refocus3} allows to integrate the obtained results with respect to $\br_b$
\begin{equation}\label{refocused3}
\Sigma_{\mathrm{ref}} (\br_a) = \int d^2 \br_b \, \Gamma_{\mathrm{ref}}(\br_a,\br_b) \sim \left|A \left( - \frac{\br_a}{M} \right) \right|^4 .
\end{equation}
to get the final high-SNR refocused image.

\section{Correlation plenoptic imaging between arbitrary planes}
\label{sec:cnp}

\begin{figure}
\centering
\includegraphics[width=0.75\textwidth]{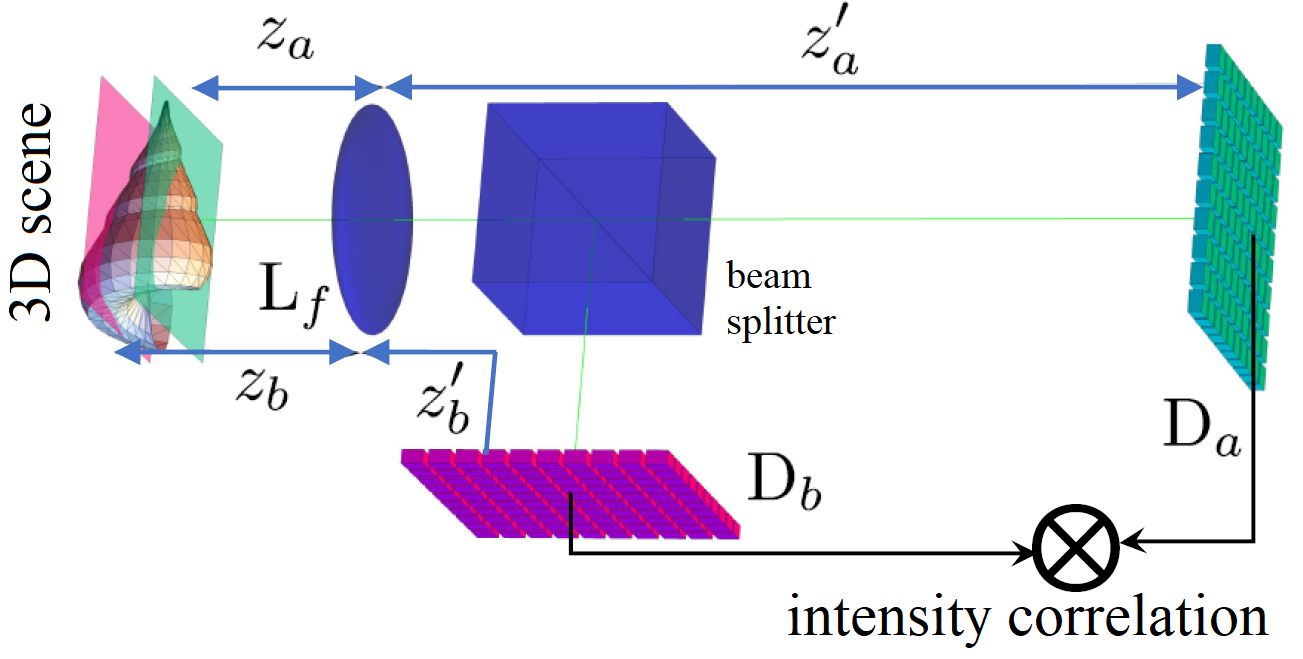}
\caption{Schematic representation of the CPI-AP protocol based on chaotic light correlations. The object is modeled as a chaotic light emitter. The lens $L_f$ generates the images of the two reference planes $\text{D}^o_a$ on the detector $\mathrm{D}_a$, and $\text{D}^o_b$ on the detector $\mathrm{D}_b$. The reference planes are arbitrarily chosen in proximity of the three-dimensional object. Combined information on the light spatial distribution and direction is obtained by computing correlations of intensity fluctuations between each pair of pixels of the two sensors.}
\label{fig:cnp:setup}
\end{figure}

The idea of correlation plenoptic imaging between arbitrary planes (CPI-AP), whose embodiment based on the correlations of chaotic light is represented in Fig.~\ref{fig:cnp:setup}, was prompted by a property of the first experimentally realized CPI device \cite{PhysRevLett.119.243602}. In that case, plenoptic imaging was based on reconstructing the direction of light between two reference planes, that could be both imaged with diffraction-limited resolution: the object plane and the source planes. A CPI-AP device is design so that the two reference planes are both placed inside the scene of interest. Besides ensuring the possibility to reconstruct two planes of the scene, instead of one, with diffraction-limited resolution, the protocol provides interesting advantages from the point of view of the depth-of-field extension, and the related setup is characterized by an outstanding compactness, compared to other CPI devices, which makes it suitable for the development of a correlation plenoptic camera.

In the scheme reported in Fig.~\ref{fig:cnp:setup}, Light from the object passes through the lens $L_f$, of focal length $f$, and is then separated by a beam splitter, and each outgoing beam is detected by one of the two spatially-resolving sensors $\mathrm{D}_a$ and $\mathrm{D}_b$. The detectors are placed in the conjugate planes of arbitrarily chosen planes $\mathrm{D}^o_a$ and $\mathrm{D}^o_b$, respectively, indicated in Fig.~\ref{fig:cnp:setup} by the same color as their conjugate sensors. The distances $z_a'$ and $z_b'$ between the lens $L_f$ and the two sensors are related by the thin-lens equations 
\begin{equation}\label{eq:cnp:tl}
\frac{1}{z_j} + \frac{1}{z_j'} = \frac{1}{f}
\qquad \text{with } j= a,b
\end{equation}
with the distances $z_a$ and $z_b$ of the respective conjugate planes from $L_f$. Also in this case, plenoptic information is contained in the correlation function \eqref{Gamma}, determined by the intensity fluctuations retrieved by the two sensors. To simplify the discussion, we consider also in this case a planar object, placed at a distance $z$ from $L_f$. Since the technique is essentially aimed at macroscopic objects, we assume in this case negligible transverse coherence: therefore, the properties of optical propagations are determined only by the intensity profile $\A(\br_o)$ of light on the object. The correlation function reads, up to irrelevant factors,
\begin{equation}\label{eq:cnp:gamma}
\Gamma(\bm\rho_a,\bm\rho_b)  = 
\left|\int d^2\bm\rho_o \A(\bm\rho_o) \int d^2\bm{\rho}_{\ell} P^*(\bm\rho_{\ell}) \int d^2\bm{\rho'}_{\ell} P(\bm\rho'_{\ell}) e^{-i k [ \phi_a(\bm\rho_o,\bm\rho_{\ell},\bm\rho_a) - \phi_b(\bm\rho_o,\bm\rho'_{\ell},\bm\rho_b) ]} \right|^2
\end{equation}
where
\begin{equation}\label{eq:cnp:phi}
\phi_j (\bm\rho_o,\bm\rho_{\ell},\bm\rho_j) =
\left(\frac{1}{z}-\frac{1}{z_j}\right)\frac{\bm\rho_{\ell}^2}{2}-
\left(\frac{\bm\rho_o}{z}+\frac{\bm\rho_j}{M_j z_j}\right) \cdot\bm\rho_{\ell} ,
\end{equation}
with $P(\br_{\ell})$ the lens pupil function and $M_j=z_j'/z_j$ the absolute magnifications of planes $\text{D}^o_j$ on sensors $\text{D}_j$, with $j=a,b$.

To clarify the imaging properties of the correlation function $\Gamma(\br_a,\br_b)$, we shall consider the geometrical-optics limit $k\to\infty$, in which the most relevant contribution to the integral in Eq.~\eqref{eq:cnp:gamma} can be estimated by applying the method of stationary phase. Actually, the stationary points of the phase $k(\phi_a-\phi_b)$, appearing in Eq.~\eqref{eq:cnp:gamma},
with respect to $\bm\rho_{\ell}$, $\bm\rho_{\ell'}$ and $\bm\rho_o$ enable us to determine, as in the case of CPM, the geometrical correspondence between points on the object and points on the sensors $\text{D}_a$ and $\text{D}_b$, providing the the dominant asymptotic contribution to the correlation function:
\begin{equation}\label{eq:gamma_geom_c}
\Gamma(\bm\rho_a,\bm\rho_b) \sim
\A^2 \left[
\frac{1}{z_b-z_a}\left(
\frac{z_a-z}{M_b}\bm\rho_b - \frac{z_b-z}{M_a}\bm\rho_a
\right)\right]  \left| P\left[
\frac{1}{z_b-z_a}\left(
- \frac{z_b}{M_a}\bm\rho_a + \frac{z_a}{M_b}\bm\rho_b
\right)\right]\right|^4 .
\end{equation}
This result shows that, independent of the distance $z$ of the object mask from the lens $L_f$, in the geometrical limit, the correlation of intensity fluctuations encodes an image of both the (squared) object intensity profile $\A^2$ and the lens pupil function $P$. 
The image of the object depends only on the coordinate of one detector, either $\br_a$ or $\br_b$, only if the object mask lies in either one of the planes $\text{D}^o_a$ or $\text{D}^o_b$, respectively. For $z=z_a$ ($z=z_b$), $\A^2$ does not depend any longer on $\bm\rho_b$ ($\bm{\rho}_a$), and the integration of the correlation function on $\bm\rho_b$ ($\bm{\rho}_a$) provides a focused image of the object:
\begin{align}\label{eq:cnp:sigmaa}
\Sigma_a(\bm\rho_a) = \int d^2\bm\rho_b\Gamma (\bm\rho_a,\bm\rho_b) \sim \A^2 \left( - \frac{\br_a}{M_a} \right) \qquad \text{if } z=z_a, \\ \Sigma_b(\bm\rho_b) = \int d^2\bm\rho_a\Gamma (\bm\rho_a,\bm\rho_b) \sim \A^2 \left( - \frac{\br_b}{M_b} \right)  \qquad \text{if } z=z_b.
\end{align}
By working in the wave optics regime, one would find that this image has the same point-spread function and depth of field as the corresponding conventional image retrived by sensor $\mathrm{D}_a$ ($\mathrm{D}_b$) alone. In the more general case in which the object does not lie in either one of the conjugate planes of the detectors and is outside the depth of field, the integral of the correlation function $\Gamma$ on either one of the detector coordinates gives rise to blurred images. 

In order to decouple the image of the object from the image of the lens and perform refocusing, we define the following linear combinations of the detector coordinates $\br_a$ and $\br_b$:
\begin{equation}
\label{eq:cnp:rhors}
\bm\rho_r = \frac{1}{z_b-z_a}\left(
\frac{z_a-z}{M_b}\bm\rho_b - \frac{z_b-z}{M_a}\bm\rho_a
\right) , \quad
\bm\rho_s = \frac{1}{z_b-z_a}\left(
- \frac{z_b}{M_a}\bm\rho_a + \frac{z_a}{M_b}\bm\rho_b
\right) .
\end{equation}
By inverting the transformation in Eq.~\eqref{eq:cnp:rhors}, we obtain the refocused correlation function:
\begin{equation}\label{eq:cnp:gammaref}
\Gamma_\text{ref}(\bm\rho_r,\bm\rho_s) =
\Gamma\left(
M_a \frac{z_a-z}{z} \bm\rho_s - M_a \frac{z_a}{z} \bm\rho_r ,
M_b \frac{z_b-z}{z} \bm\rho_s - M_b \frac{z_b}{z} \bm\rho_r
\right) \sim 
\A\left(\bm\rho_r\right)^2
\left| P\left(\bm\rho_s \right)\right|^4
\end{equation}
From the last line of Eq.~\eqref{eq:cnp:rhors}, it is evident that the performed linear transformation of the argument of $\Gamma$ realigns all the displaced images corresponding to different values of $\br_b$, and the refocused correlation function obtained by ``reordering'' the correlation function. No blurring occurs anymore upon integrating the refocused correlation function over the variable $\bm\rho_s$; in fact, this integral gives the final refocused image
\begin{equation}\label{eq:cnp:sigmar}
\Sigma_\text{ref}(\bm\rho_r) =
\int d^2\bm\rho_s \Gamma_\text{ref}(\bm\rho_r,\bm\rho_s) \sim  \A\left(\bm\rho_r\right)^2 .
\end{equation}
Although any combination $\br_s$ gives a focused image of the object, the integration over $\br_s$ reported  in Eq.~\eqref{eq:cnp:sigmar} enables to exploit the whole signal collected by the two detectors, hence, to considerably increase the signal-to-noise ratio of the final image.

\section{CONCLUSIONS}

We reviewed the main structural features and the working principles of correlation plenoptic microscopy and correlation plenoptic imaging between arbitrary planes. Though stemming from the same original idea of CPI, the two methods are characterized by specificities that make them particularly suitable for microscopy and photography, respectively. Both protocols open the possibility to achieve combinations of resolution and depth of field that are unattainable by both standard imaging and first-order plenoptic imaging\cite{CPM,CPI-AP}. Future research will be devoted to the speedup of the described techniques, in order to make the acquisition times comparable to those of state-of-the-art 3D imaging devices. A relevant point along this path is represented by an accurate evaluation of the signal-to-noise properties of the correlation function (already studied in previous CPI devices\cite{scala}. It is also worth noticing that we are working towards CPI-AP with entangled-photons illumination, as it would represent an interesting configuration to exploit the improvement in signal-to-noise ratio that entangled photon pairs are known
to allow in the low-photon-flux regime\cite{Samantaray,subshot}.

\section*{ACNOWLEDGMENTS}
This work is supported 1) by the Italian Istituto Nazionale di Fisica Nucleare, the Swiss National Science Foundation, the Greek General Secretariat for Research and Technology, the Czech Ministry of Education, Youth and Sports, under the QuantERA programme (Qu3D project), which has received funding from the European Union’s Horizon 2020 research and innovation programme, 2) by Istituto Nazionale di Fisica Nucleare projects PICS4ME and INTEFF-TOPMICRO, and 3) by PON ARS $01\_00141$ ``CLOSE -- Close to Earth'' of Ministero dell'Istruzione, dell'Universit\`a e della ricerca (MIUR).

%\bibliography{report}

\begin{thebibliography}{10}

\bibitem{PI}
E.~H. {Adelson} and J.~Y.~A. {Wang}, ``Single lens stereo with a plenoptic
  camera,'' {\em IEEE Trans. Pattern Anal. Mach.
  Intell.}~{\bf 14}, 99 (1992).

\bibitem{PI1}
R.~NG, M.~{Levoy}, M.~{Br{\'e}dif}, G.~{Duval}, M.~{Horowitz}, and Pat{
  Hanrahan}, ``Light field photography with a hand-held plenoptic camera,''
  {\em Stanford University Computer Science Tech Report CSTR 2005-02} (2005).

\bibitem{Broxton:13}
M.~Broxton, L.~Grosenick, S.~Yang, N.~Cohen, A.~Andalman, K.~Deisseroth, and
  M.~Levoy, ``Wave optics theory and 3-d deconvolution for the light field
  microscope,'' {\em Opt. Express}~{\bf 21}, 25418 (2013).

\bibitem{Xiao:13}
X.~Xiao, B.~Javidi, M.~Martinez-Corral, and A.~Stern, ``Advances in
  three-dimensional integral imaging: sensing, display, and applications,''
  {\em Appl. Opt.}~{\bf 52}, 546 (2013).

\bibitem{PI6}
R.~Prevedel, Y.-G. Yoon, M.~Hoffmann, N.~Pak, G.~Wetzstein, S.~Kato,
  T.~Schr{\"o}del, R.~Raskar, M.~Zimmer, E.~S. Boyden, and A.~Vaziri,
  ``Simultaneous whole-animal 3d imaging of neuronal activity using light-field
  microscopy,'' {\em Nat. Methods}~{\bf 11}, 727 (2014).

\bibitem{c72b58c6767e455488b7dd420d55b06a}
V.~Adhikarla, J.~Sodnik, P.~Szolgay, and G.~Jakus, ``Exploring direct 3d
  interaction for full horizontal parallax light field displays using leap
  motion controller,'' {\em Sensors}~{\bf 15}, 8642 (2015).

\bibitem{PI2}
R.~Ng, ``Fourier slice photography,'' {\em ACM Trans. Graph.}~{\bf 24},
  735 (2005).

\bibitem{PI3}
T.~Georgiev and A.~Lumsdaine, ``{The multifocus plenoptic camera},'' in {\em
  Digital Photography VIII}, edited by S.~Battiato, B.~G. Rodricks, N.~Sampat, F.~H.
  Imai, and F.~Xiao, {\bf 8299}, pp.~69 -- 79 (SPIE International Society for
  Optics and Photonics, 2012).

\bibitem{PI4}
B.~Goldluecke, O.~Klehm, S.~Wanner, and E.~Eisemann, {\em Plenoptic Cameras},
  ch.~Digital Representations of the Real World: How to Capture, Model, and
  Render Visual Reality, pp.~65--77 (CRC Press, 2015).

\bibitem{5989827}
T.~E. {Bishop} and P.~{Favaro}, ``The light field camera: Extended depth of
  field, aliasing, and superresolution,'' {\em IEEE Transactions on Pattern
  Analysis and Machine Intelligence}~{\bf 34}, 972 (2012).

\bibitem{PI5}
M.~Broxton, L.~Grosenick, S.~Yang, N.~Cohen, A.~Andalman, K.~Deisseroth, and
  M.~Levoy, ``Wave optics theory and 3-d deconvolution for the light field
  microscope,'' {\em Opt. Express}~{\bf 21}, 25418 (2013).

\bibitem{PI7}
K.~C. Zheng, B.~Curless, D.~Salesin, S.~K. Nayar, and C.~Intwala,
  ``Spatio-angular resolution tradeoffs in integral photography,'' in \textit{Eurographics Symposium on
Rendering, 2006}, edited by T. Akenine-Möller, and W.
Heidrich (The Eurographics Association, Geneva, 2006).

\bibitem{firstCPI}
M.~D'Angelo, F.~V. Pepe, A.~Garuccio, and G.~Scarcelli, ``Correlation plenoptic
  imaging,'' {\em Phys. Rev. Lett.}~{\bf 116}, 223602 (2016).

\bibitem{PhysRevLett.119.243602}
F.~V. Pepe, F.~Di~Lena, A.~Mazzilli, E.~Edrei, A.~Garuccio, G.~Scarcelli, and
  M.~D'Angelo, ``Diffraction-limited plenoptic imaging with correlated light,''
  {\em Phys. Rev. Lett.}~{\bf 119}, 243602 (2017).

\bibitem{CPIent}
F. V.~Pepe, F.~Di~Lena, A.~Garuccio, G.~Scarcelli, and M.~D'Angelo, ``Correlation
  plenoptic imaging with entangled photons,'' {\em Technologies}~{\bf 4}, 17
  (2016).

\bibitem{overview}
F.~Di~Lena, F. V.~Pepe, A.~Garuccio, and M.~D'Angelo, ``Correlation plenoptic
  imaging: An overview,'' {\em Appl. Sci.}~{\bf 8}, 1958 (2018).

\bibitem{article}
F.~Pepe, O.~Vaccarelli, A.~Garuccio, G.~Scarcelli, and M.~D'Angelo, ``Exploring
  plenoptic properties of correlation imaging with chaotic light,'' {\em
  J. Opt.}~{\bf 19}, 114001 (2017).
  
\bibitem{CPM}
A.~Scagliola, F.~{Di Lena}, A.~Garuccio, M.~D'Angelo, and F.~V. Pepe,
  ``Correlation plenoptic imaging for microscopy applications,'' {\em Phys.
  Lett. A}~{\bf 384}, 126472 (2020).

\bibitem{CPI-AP}
F. Di Lena, G. Massaro, A. Lupo, A. Garuccio, F. V. Pepe, and M. D'Angelo, ``Correlation Plenoptic Imaging between Arbitrary Planes,'' {\em Opt. Express}~\textbf{28}, 35857 (2020).

\bibitem{pittman}
T.~Pittman, D.~V.~Strekalov, D.~N.~Klyshko, M.~Rubin, A.~Sergienko, and
  Y.~Shih, ``Two-photon geometric optics,'' {\em Phys. Rev. A}~{\bf 53},
  2804 (1996).

\bibitem{ghost}
B.~I.~Erkmen and J.~H.~Shapiro, ``Ghost imaging: From quantum to classical to
  computational,'' {\em Adv. Opt. Photonics}~{\bf 2},
  405 (2010).

\bibitem{gatti}
A.~Gatti, E.~Brambilla, M.~Bache, and L.~A. Lugiato, ``Ghost imaging with thermal light: comparing entanglement and classical correlation,'' \textit{Phys. Rev. Lett.}~\textbf{93}, 093602 (2004).

\bibitem{laserphys}
M.~D'Angelo and Y.~Shih, ``Quantum imaging,'' \textit{Laser Phys. Lett.}~\textbf{2}, 567 (2005).

\bibitem{valencia}
A.~Valencia, G.~Scarcelli, M.~D’Angelo, and Y.~Shih, ``Two-photon imaging with thermal light,'' \textit{Phys. Rev. Lett.} \textbf{94}, 063601 (2005).

\bibitem{scarcelliPRL}
G.~Scarcelli, V.~Berardi, and Y.~Shih, ``Can two-photon correlation of chaotic light be considered as correlation of intensity fluctuations?,'' Physical Review Letters \textbf{96}, 063602 (2006).

\bibitem{mandel}
L. Mandel, E. Wolf, {\em Optical Coherence and Quantum Optics} (Cambridge University Press, Cambridge, 1995).

\bibitem{scala}
G.~Scala, M.~D'Angelo, A.~Garuccio, S.~Pascazio, and F. V.~Pepe, ``Signal-to-noise
  properties of correlation plenoptic imaging with chaotic light,'' {\em
  Phys. Rev. A}~{\bf 99}, 053808 (2019).



\bibitem{Samantaray}
N.~Samantaray, I.~Ruo-Berchera, A.~Meda, and M.~Genovese, ``Realization of the
  first sub-shot-noise wide field microscope,'' {\em Light Sci.
  Appl.}~{\bf 6}, e17005 (2017).

\bibitem{subshot}
G.~Brida, M.~Genovese, and I.~Ruo~Berchera, ``Experimental realization of
  sub-shot-noise quantum imaging,'' {\em Nat. Photonics}~{\bf 4}, 227 (2010).

\end{thebibliography}

\end{document}